\documentstyle[12pt]{article}
\begin{document}

\begin{center}

{Classification of states  of single-$j$ fermions with  $J$-pairing interaction  }

{ Y. M. Zhao$^{a,b,c}$ and     A. Arima$^{d}$}

{
$^a$  Cyclotron Center,  Institute of Physical Chemical Research (RIKEN), \\
Hirosawa 2-1,Wako-shi,  Saitama 351-0198,  Japan \\
$^b$ Department of Physics, Saitama University, Saitama-shi,
Saitama 338 Japan \\
$^c$ Department of Physics,  Southeast University, Nanjing 210018 China \\
$^d$ The House of Councilors, 2-1-1 Nagatacho,
Chiyodaku, Tokyo 100-8962, Japan }

\date{\today}

\end{center}

\vspace{0.3in}

\begin{small}
In this paper we  show that a system of
three fermions  is exactly solvable for the case of  a single-$j$
 in the presence of an angular momentum-$J$  pairing interaction.
On the basis of the solutions for this system, we obtain  new sum rules
for six-$j$ symbols.
When the Hamiltonian contains only an interaction
between pairs of fermions coupled to spin $J=J_{\rm max}=2j-1$, 
the ``non-integer" eigenvalues  of three fermions
with angular momentum $I$ around the maximum
appear as ``non-integer" eigenvalues of four fermions 
if $I$ is around (or
larger than) $J_{\rm max}$. 
This pattern is also found  in five and six fermion systems.
A boson system with spin $l$ exhibits a similar pattern.

\end{small}


\newpage

\section{Introduction}
To obtain a simple solution of the many-body  Schr\"odinger equation
 is a long dream of physicists. There have been
numerous efforts to obtain analytical solutions which
do not require diagonalization of the secular equation by using
computers. Among the many efforts in nuclear physics,
the Elliott Model \cite{Elliott}, the
seniority scheme \cite{Racah}, the $s$ and $d$ interacting boson model
\cite{Arima}
and a similar model by using  schematic
$S$ and $D$ pairs, the fermion dynamical symmetry
model \cite{Ginocchio} 
  are   successful examples along this line.

In Ref. \cite{nucl-th/0305095}  we showed that
 for a large array of states of  four fermions in  a 
single-$j$ shell with only the $J=J_{\rm max}$ pairing interaction,
the eigenvalues are   asymptotically integers
labeled by   numbers of $J=J_{\rm max}$ 
pairs in the wavefunction,
 and that  those  corresponding
 wavefunctions of these low $I^{(4)}$ ($I^{(4)}$ is the total
 angular momentum for a state of the four fermions)   states
 are readily constructed in the nucleon
 pair basis. Besides the ``integer" eigenvalues (as explained
 in Sec. 2), there are
eigenvalues  not  close to integers when $I^{(4)}$ is around or larger than $2j$,
and little was known about these states.  It would be desirable to discuss
both the ``integer" and ``non-integer"  eigenvalues
on a more general footing.  This is one of the goals in this paper.

In Sec. II of this paper, we shall first
study $n=3$ systems ($n$ is the  number of fermions),
which are readily solvable
for any    $J$ pairing interaction only.
The solutions of $n=3$  provide  an appropriate
platform to explain the main idea of this paper.
In Sec. III, we   report   relations between
the eigenvalues of $n=3$-, 4-, 5-, and 6-particle systems with  
the $J_{\rm max}$ pairing interaction. Using these relations
one may obtain approximate values for  both the eigenvalues and
wave functions of low-lying  states of these systems. 
We propose a hypothesis by which one readily
obtains, to  a high precision and without diagonalization
of the Hamiltonian,  the  wavefunctions of states corresponding to some 
non-integer eigenvalues discussed in our earlier work \cite{nucl-th/0305095}.
A summary and discussion is given in Sec. IV.
In the Appendix, we present a number of new sum rules for six-$j$
symbols, some of which were derived recently by    Talmi \cite{communication}.

In this paper we use a convention that
 $j$  ($j'$) is a half integer, and that $l$ ($l'$) is an
 integer.  They
correspond to the angular momenta of the single-particle levels
of fermions and spin carried by boson, respectively.
$J$ is used as the angular momentum coupled by two fermions
in a single-$j$ shell or two bosons with spin $l$; the maximum of $J$,
$J_{\rm max}$,  $= 2j-1$ for fermions and $2l$ for two bosons.
We use superscript $( n )$ to specify the particle  number $n$ in 
the angular momentum $I^{(n)}$ and 
the eigenvalue $E^{(n)}_{I^{(n)}, J(j)}$.

\section{Three fermions in the presence of $H_J$ only}

The  pairing interaction   which couples two fermions
 to an angular momentum $J$ is
\begin{eqnarray}
&&  H_J =
G_J \sum_{M = -J}^J A_M^{J \dagger} \ A_{M}^{J }, ~ ~
A_M^{J \dagger} = \frac{1}{\sqrt{2}} \left[ a_{j}^{\dagger}
\times a_{j}^{\dagger}
     \right]^{J},
\nonumber \\
&&
     {A}_M^{J} = - (-1)^M\frac{1}{\sqrt{2}} \left[ \tilde{a}_{j} \times
     \tilde{a}_{j} \right]_{-M}^J, ~ ~ ~
     \tilde{A}^J = - \frac{1}{\sqrt{2}} \left[ \tilde{a}_{j} \times
     \tilde{a}_{j} \right]^{J}.   \label{pair}
\label{H}
\end{eqnarray}
where $[~]_M^{J}$ means coupled to angular momentum $J$ and projection $M$.
We take $G_J=-1$ in this paper.

We now consider the pair basis of three nucleons
\begin{equation}
|j^3 [j J]  I^{(3)}, M \rangle =
     \frac{1}{\sqrt{ N^{I^{(3)}}_{j J; j J}}}
     \left(  a^{\dag}_j  \times
A^{J \dag} \right)^{I^{(3)}}_M |0 \rangle, \label{basis0}
\end{equation}
where $N^{I^{(3)}}_{j J; j J} $ is the diagonal matrix element of the
normalization matrix
\begin{eqnarray}
 N^{I^{(3)}}_{j J^{\prime}; j J} = \langle 0 |\left(  a_j
\times A^{J'} \right)^{I^{(3)}}_M  \left(  a_j^{\dag}  \times
A^{J\dag} \right)^{I^{(3)}}_M |0 \rangle.
\end{eqnarray}
In general this basis is over complete and the normalization matrix
may have zero eigenvalues for a given $I^{(3)}$.

We first rewrite  the
matrix elements of $H_J$  and  $N^{I^{(3)}}_{j J^{\prime}; j J}$ as follows
\cite{Yoshinaga}.
\begin{eqnarray}
&&      \langle j^3 [j  K^{\prime}] I^{(3)},M |
H_J  |j^3 [j K] I^{(3)},M\rangle     \nonumber \\
&&= -\frac {1}{ \sqrt{ N^{I^{(3)}}_{j K^{\prime}; j K^{\prime}} }
\sqrt{ N^{I^{(3)}}_{j K; j K} }  }
\sum_L (-)^{I^{(3)}+J-L} \frac{\hat{L}}{2I^{(3)}+1} \times  \nonumber \\
&&
\langle 0 |  \left(
\left[
     \left(  \tilde{a}_j  \times
\tilde{A}^{K'} \right)^{I^{(3)}}, A^{J \dag}   \right]^{L} \times
\left[ \tilde{A}^{J},
     \left(  a^{\dag}_j  \times
A^{K \dag} \right)^{I^{(3)}}    \right]^{L}
              \right)^{(0)}
 |0 \rangle ~,           \nonumber \\
&& N^{I^{(3)}}_{j J^{\prime}; j J} = \frac{1}{\hat{I}^{(3)}}
\langle 0 | \left( \tilde{a}_j \times \left[ \tilde{A}^{J'},
\left(  a_j^{\dag} \times A^{J \dag} \right)^{I^{(3)}} \right]^{j}  \right)
|0 \rangle,
\label{to_be_calculate}
\end{eqnarray}
where $\hat{L}$ is a short hand notation of $\sqrt{2L+1}$.

According to  Eq. (12a) of Ref. \cite{Chen1} and
 Eqs. (3.10a) and (3.10b) \cite{Chen2},
\begin{eqnarray}
&&
\left[ \tilde{a}_j,
   \left(a_j^{\dag} \times \tilde{a}_j \right)^t   \right]^I
= -\delta_{I^{(3)}, j} (-)^{t} \frac{\hat{t} }{\hat{j} } \tilde{a_j}.
\nonumber \\
&& \left[\tilde{A}^r, A^{s\dag} \right]^t =
\hat{r} \delta_{r, s} \delta_{t, 0} -
 2 \hat{r} \hat{s}
    \left\{ \begin{array}{ccc}
    r    & s  & t \\
    j    & j  & j  \end{array} \right\}
\left(a_j^{\dag} \times
 \tilde{a}_j \right)^t. \label{commutator1}
\end{eqnarray}
Using commutators in (\ref{commutator1}),
we obtain
\begin{eqnarray}
&& \langle 0 | \left[
     \left(  \tilde{a}_j  \times
\tilde{A}^{K'} \right)^{I^{(3)}}, A^{J \dag}   \right]^{L}
= \langle 0 |  \left[  (-)^{I^{(3)}-j} \delta_{L,j} \delta_{K',J}
\frac{\hat{I^{(3)}}}{\hat{j}} \tilde{a}_j
\right.  \nonumber \\
&& ~ \left. - 2 \hat{K}' \hat{J} \sum_t
(-)^{j+K'+L+J}
\hat{I^{(3)}} \hat{t}
    \left\{ \begin{array}{ccc}
    j    & K'  & I^{(3)} \\
    J    & L   & t  \end{array} \right\}
    \left\{ \begin{array}{ccc}
    K'    & J  & t \\
    j    & j  & j  \end{array} \right\}
\left( \tilde{a}_j \times
\left(a_j^{\dag} \times  \tilde{a}_j \right)^t
\right)^{(L)}  \right] \nonumber \\
&&
= (-)^{I^{(3)}-j}
\frac{\hat{I^{(3)}}}{\hat{j}} \delta_{L, j}
\left( \delta_{K', J} + 2 \hat{K}' \hat{J}
    \left\{ \begin{array}{ccc}
    K'    & j  & j \\
    J    & j  & I^{(3)}  \end{array} \right\}  \right) \langle 0 | \tilde{a}_j ~.
    \label{commutator4}
\end{eqnarray}
where a sum rule
\begin{equation}
\sum_t (-)^{t+j+I^{(3)}} (2t+1)
    \left\{ \begin{array}{ccc}
    K'    & J  & t \\
    j    & j  & j  \end{array} \right\} 
    \left\{ \begin{array}{ccc}
    K'    & J  & t \\
    j    & j  & I^{(3)}  \end{array} \right\}  =
    \left\{ \begin{array}{ccc}
    K'    & j  & j \\
    J    & j  & I^{(3)}  \end{array} \right\} 
\end{equation}
is used.
The Hermitian conjugate of Eq.~(\ref{commutator4}) yields
\begin{eqnarray}
&&   \left[ \tilde{A}^{J},
     \left(  a^{\dag}_j  \times
A^{K \dag} \right)^{I^{(3)}}    \right]^{L}  | 0 \rangle
= (-)^{2I^{(3)}+J-L-j}  \delta_{L, j}
\frac{\hat{I^{(3)}}}{\hat{j}}
\left( \delta_{K,  J} + 2 \hat{K}  \hat{J}
    \left\{ \begin{array}{ccc}
    K     & j  & j \\
    J    & j  & I^{(3)}  \end{array} \right\}  \right) {a}_j^{\dag} | 0 \rangle .
\nonumber \\
\label{commutator5}
\end{eqnarray}
Substituting Eqs. (\ref{commutator4}) and (\ref{commutator5}) into
Eq.~(\ref{to_be_calculate}), we obtain
\begin{eqnarray}
&& N^{I^{(3)}}_{j J^{\prime}; j J}
=\delta_{J', J}  + 2 \hat{J}      \hat{J^{\prime}}
     \left\{ \begin{array}{ccc}
     J    & j  & I^{(3)} \\
     J^{\prime}  & j & j   \end{array} \right\} , \nonumber \\
&&      \langle j^3 [j  K^{\prime}] I^{(3)},M |
H_J  |j^3 [j K] I^{(3)},M\rangle
= -\frac {1}{{\sqrt{N^{I^{(3)}}_{j K^{\prime}; j K^{\prime}}
N^{I^{(3)}}_{j K; j K} } }}
  N^{I^{(3)}}_{j  K^{\prime}; j J}
  N^{I^{(3)}}_{j J; j  K}  .
\label{matrix}
\end{eqnarray}

Below we  explain that there
is at most  one non-zero eigenvalue for each $I^{(3)}$ of $n=3$ with
$H=H_J$. For a fixed $J$ and for any $I^{(3)}$, we construct the 
$|j^3 J: I^{(3)} \rangle \rangle $ and other 
states $|j^3 K: I^{(3)} \rangle \rangle$ $(K \neq J)$ which are orthogonal to
$|j^3 J: I^{(3)} \rangle \rangle$ as follows. 
\begin{eqnarray}
&&      |j^3 J: I^{(3)} \rangle \rangle= |j^3 [jJ ]I^{(3)} \rangle ~ ,  \nonumber \\ 
&& |j^3 K: I^{(3)} \rangle \rangle= |j^3 [j K] I^{(3)} \rangle
- \frac
{N^{I^{(3)}}_{j K; j J}}{\sqrt{N^{I^{(3)}}_{j J; j J} N^{I^{(3)}}_{j K; j K}}}|j^3
[jJ]I^{(3)} \rangle,\ (K \neq J). \label{basis}
\end{eqnarray}

Using Eq. (\ref{matrix}), we easily confirm that all matrix elements of
the Hamiltonian in the basis (\ref {basis}),
$ \langle \langle j^3 K':I^{(3)} | H_J  | j^3 K: I^{(3)} \rangle\rangle$,  are
zero except   $ \langle \langle |j^3 [jJ ]I^{(3)} | H_J|j^3 [jJ ]I^{(3)} \rangle\rangle$
= $ N^{I^{(3)}}_{j J; j J}$.
Thus all the eigenvalues of $n = 3$ for a given $I^{(3)}$
 are zero for $H=H_J$ except for the
state with one pair with spin  $J$, with an  eigenvalue $E_{I^{(3)},J(j)}^{(3)}$
(the number in superscript specify the particle number $n$)  given by
$- N^{I^{(3)}}_{jJ; jJ}$.  This result was also proved
recently by Talmi in terms of coefficients of fractional
parentage \cite{communication}.

Next we explain why the  eigenvalues for the
$n=3$ cases are close to integers
when $H=H_{J_{\rm max}}$.  As shown above,
the wave function of the lowest energy state for each $I^{(3)}$ is given
by $\left( \left(a_j^{\dag} \times a_j^{\dag} \right)^{(J_{\rm max})} \times
a_j^{\dag} \right)^{I} |0 \rangle$. The
eigenvalue $E_{I^{(3)},J(j)}^{(3)}$   equals to $-1$ subtracted by
a six-$j$ symbol (refer to Eq. (\ref{matrix})),
this six-$j$ symbol  is in fact 
very close to zero when $I^{(3)}$ is not close to $I^{(3)}_{\rm max}=3j-3$.
The lowest eigenvalue for each $I^{(3)}$ $(I^{(3)} \ge  j-1$) is thus
 very close to $-1$  unless $I^{(3)}\sim I^{(3)}_{\rm max}$.
All eigenvalues for $I^{(3)} \le j-2$ are zero.
To show that the six-$j$ symbols
involved in Eq. (\ref{matrix})  asymptotically vanishes, we
list  a few formulas of these six-$j$ symbols:
\begin{eqnarray}
&&    \left\{ \begin{array}{ccc}
    j    & j-1  & 2j-1 \\
    j    & j  & 2j-1  \end{array} \right\} =
    -\frac{2 j (2j-1)!}{ (4j-1)!} ~; \nonumber \\
&&    \left\{ \begin{array}{ccc}
    j    & j  & 2j-1 \\
    j    & j  & 2j-1  \end{array} \right\} =
    \frac{ j (4j-3) (2j-1)!}{ (4j-1)!} ~; \nonumber \\
&&    \left\{ \begin{array}{ccc}
    j    & j+1  & 2j-1 \\
    j    & j  & 2j-1  \end{array} \right\} =
   - \frac{ j (4j^2-4j-1) (2j-1)!}{ (4j-1)!} ~; \nonumber \\
\end{eqnarray}
which are less than $10^{-14}$ in magnitude for $j=31/2$. Clearly,
the approximate integer eigenvalues of $n=3$ with $H=H_{J_{\rm max}}$
comes from the fact that the six-$j$ symbol
$   \left\{ \begin{array}{ccc}
    j    & I^{(3)}  & 2j-1 \\
    j    & j  & 2j-1  \end{array} \right\}$   are negligible
    unless $I^{(3)} \sim I^{(3)}_{\rm max}$.

The ``non-integer" eigenvalues with $I^{(3)}\sim I^{(3)}_{\rm max}$ are also
readily obtained:
\begin{eqnarray}
&& -E^{(3)}_{I^{(3)}_{\rm max}, J_{\rm max}(j)} = \frac{9}{4} + \frac{3}{4(4j-3)}; \nonumber \\
&& -E^{(3)}_{I^{(3)}_{\rm max}-2, J_{\rm max}(j)} = \frac{27}{16} - \frac{15}{32(4j-5)} -
 \frac{21}{32(4j-3)}; \nonumber \\
&& -E^{(3)}_{I^{(3)}_{\rm max}-3, J_{\rm max}(j)} = \frac{9}{16} + \frac{15}{32(4j-5)} +
 \frac{45}{32(4j-3)}; \nonumber \\
&& -E^{(3)}_{I^{(3)}_{\rm max}-4, J_{\rm max}(j)} = \frac{81}{64} + \frac{105}{512(4j-7)} -
 \frac{15}{256(4j-5)} -  \frac{1155}{512(4j-3)} ; \nonumber \\
&& -E^{(3)}_{I^{(3)}_{\rm max}-5, J_{\rm max}(j)} = \frac{27}{32} - \frac{105}{256(4j-7)} -
 \frac{105}{128(4j-5)} + \frac{819}{256(4j-3)} ; \nonumber \\
&& -E^{(3)}_{I^{(3)}_{\rm max}-6, J_{\rm max}(j)} = \frac{279}{256} - \frac{315}{4096(4j-9)} +
 \frac{1785}{4096(4j-7)} \nonumber \\
&& ~~~~~~~ ~~~~~~~~~ ~ +
 \frac{9135}{4096(4j-5)} - \frac{17325}{4096(4j-3)} ; \nonumber \\
&& -E^{(3)}_{I^{(3)}_{\rm max}-7, J_{\rm max}(j)} = \frac{243}{256} + \frac{945}{4096(4j-9)} -
 \frac{315}{4096(4j-7)} \nonumber \\
&& ~~~~~~~ ~~~~~~~~~ ~ -
 \frac{17325}{4096(4j-5)} + \frac{21879}{4096(4j-3)} ; \nonumber \\
&& -E^{(3)}_{I^{(3)}_{\rm max}-8, J_{\rm max}(j)} = \frac{1053}{1024} + \frac{3465}{131072(4j-11)}
- \frac{12915}{32768(4j-9)}  \nonumber \\
&& ~~~~~~~ ~~~~~~~~~ ~
-  \frac{58905}{65536(4j-7)}  +
 \frac{225225}{32768(4j-5)} - \frac{855855}{131072(4j-3)} ; \nonumber \\
&& -E^{(3)}_{I^{(3)}_{\rm max}-9, J_{\rm max}(j)} = \frac{63}{64} - \frac{3465}{32768(4j-11)}
+ \frac{3465}{8192(4j-9)}  \nonumber \\
&& ~~~~~~~ ~~~~~~~~~ ~
+  \frac{45045}{16384(4j-7)} -
 \frac{83655}{8192(4j-5)} + \frac{255255}{32768(4j-3)} ;
\label{non-integer}
\end{eqnarray}
etc.
We see that these above eigenvalues $E^{(3)}_{I^{(3)}, J_{\rm max}(j)}$ stagger and saturate 
at $-1$ as $I^{(3)}$ becomes smaller and smaller (but $I^{(3)}\ge j-1$).
In the large $j$ limit, the non-zero eigenvalue for each $I^{(3)}$
takes  the first term;
for a very small $j$ value, e.g., $j=9/2$, the
  eigenvalue $E_{ I^{(3)}_{\rm max}-7, J_{\rm max} (9/2)}$
  (i.e., $I^{(3)}=7/2$, 
   $E^{(3)}_{7/2, J_{\rm max}(9/2) }$   equals to  $-\frac{714}{715}$)
is already very close to $-1$ (within a precision of $10^{-2}$).
This explains why we frequently obtain 
asymptotic $-1$ eigenvalues for $H=H_{J_{\rm max}}$
and $n=3$.
For a state which has $E^{(3)}_{I^{(3)}, J_{\rm max}(j)} \sim -1$, the
corresponding wavefunction can be
understood as a single-$j$ ``spectator" coupled to one pair with spin
$J_{\rm max}$.
There is no such a spectator for $I^{(3)}\sim I^{(3)}_{\rm max}$ states, although
their wave functions can be written as $\ j^2 (J_{\rm max}) j: I^{(3)} \rangle$. 

We note that three bosons with spin $l$ exhibit a similar
pattern: there is up to one non-integer eigenvalue
for each $I$  in the presence of boson Hamiltonian $H_J$,
\begin{equation}
E^{(3)}_{I^{(3)}, J(l)} = - 1 - 2(2J+1)
     \left\{ \begin{array}{ccc}
     J    & l  & I^{(3)} \\
     J    & l & l . \end{array} \right\} ~ ,
\end{equation}
i.e., this non-zero (zero) eigenstate are given by a pair
with spin $J$, i.e., $\left(b_l^{\dag} \times b_l^{\dag} \right)^{(J)}$,
coupled with a single boson operator $b_l^{\dag}$.

\section{Relations between  states of $n=3$ and 4
for $H=H_{J_{\rm max}}$. }

 In this section, we
first discuss the cases with $n=4$. In Ref. \cite{nucl-th/0305095},   it was
found that the eigenvalues of $n=4$ are asymptotically  0, $-1$ or $-2$ for
small $I^{(4)}$. These states are constructed by coupling one or two pairs with spin
$J=J_{\rm max}$. However, some ``non-integer" eigenvalues appear as  $I^{(4)}$ is
larger than $2j-9$. These values are very stable for $2j-8 \le 4j-12$, and the
origin for these states was unknown.

Let us compare the eigenvalues of a system with $n=3$ and $n=4$ fermions with
$H=H_{J_{\rm max}}$  for $j=31/2$. The distribution  of all non-zero
eigenvalues for $n=3$ and 4 is plotted in Fig. 1(a)-(c)     \footnote{The inset
in Fig. 1(b) is re-scaled in order to see more clearly the exceptions of
energies which are  not close to those of $n=3$.}, where   (a), (b) and (c)
correspond  to the range of $|E|$ from 0 to 0.8, 0.8 to 1.5, 1.5 to 3.8,
respectively.

From Fig. 1, we see that these eigenvalues are  clustered at  a few values
but with exceptions. The ``clustered" values are {\it very} close to the
eigenvalues of $n=3$ \footnote{There is one peak at $2.0$ which was explained
using two pairs with spin $J=2j-1$ in Ref. \cite{nucl-th/0305095}. }.  This
indicates that the eigenstates of $n=4$ are closely related to those of $n=3$.

For $j=31/2$ and $n=4$ the total number of states
is 790. The number of states with non-zero eigenvalues
is 380. Within a precision $10^{-2}$,
the eigenvalues of these 308 states
are located at the eigenvalues of $n=3$,
and 21 states have eigenvalues closely at $-2$. 
We note that almost all the ``non-integer" eigenvalues
of $n=4$ can be rather accurately given by 
one of three-particle clusters with $I^{(3)} \sim I^{(3)}_{\rm max}$
coupled to  a single-$j$ particle.  In this example only 
four states with $I^{(4)}=48$,   two states with $I^{(4)}=46$, and two states
with $I^{(4)}=44$ cannot be understood by either  
one of  three-particle clusters with $I^{(3)} \sim I^{(3)}_{\rm max}$ 
 coupled to a single-$j$ particle 
or two pairs with one or two spins being $J_{\rm max}$.

For example,  the peak  for $n=4$ in Fig. 1(c)
 near $2.25$  is very close to
the energy of $|E_{I^{(3)}_{\rm max}, J_{\rm max}(j)}^{(3)}|$ of $n=3$. For $j=31/2$, the maximum
angular momentum $I_{\rm max}^{(3)}$ of three fermions is $\frac{87}{2}$. The
$E_{I^{(3)}_{\rm max}, J_{\rm max}(j)}^{(3)}$ =$-\frac{267}{118}$=$-$2.26271186440677966.
The minimum $I^{(4)}$ obtained by coupling a three-body cluster 
 with $I^{(3)}=I^{(3)}_{\rm max}$
to a single-$j$ particle is given by $3j-3-j=2j-3$ (the triangle
relation for vector couplings)  and here 28. We find  that the lowest
eigenvalue  of $I^{(4)}=28$ for $n=4$ obtained from a shell model
diagonalization is $-2.26271186440689$. The
$E_{I^{(4)}_{\rm max}, J_{\rm max}(j)}^{(4)}$
(close to $-2.26$)   with $I^{(4)}$ 
between  $28$ to 56 are listed in Table I. Two observations can be made: (1)
the lowest state of each $I^{(4)}$ are well separated from the second lowest one, and
(2), there is {\it no} eigenvalue which is smaller than -2.0 when $I^{(4)}$ is lower
than $2j-3$  for $n=4$ 
\footnote{
We also note
 that the above nearly equality is  {\it asymptotic} for
 a rather large $j$, not exact.
For examples, when $j$ is very small,
for $j=7/2$, the
 energy of $E_{I^{(3)}_{\rm max}, J_{\rm max}(j)}^{(3)}$
 is $-\frac{51}{22}$=2.31818182 while the lowest 
energies of $I^{(4)}=2j-3$=4 for four fermions
obtained by diagonalization is $-\frac{8}{3}=2.66666667$;
for  $j=9/2$,
the  energy of $E_{I^{(3)}_{\rm max}, J_{\rm max}(j)}^{(3)}$
 is $-\frac{23}{10}$ while the lowest 
energies of $I^{(4)}=2j-3$=6 for four fermions 
obtained by diagonalization is $-2.34965034965038$.}.

We also see that the overlap of the wave function obtained by the exact shell
model with the state constructed as
 the $I_{\rm max}^{(3)}$ state  coupled to
 a single-$j$ particle, $[I_{\rm max}^{(3)}\times j]^{I^{(4)}}$,
 is very close to 1.
This may be argued as follows. The eigenvalue for the  lowest  state of $n=4$
for $I^{(4)} \ge 28$ is {\it very} close to the matrix element of the Hamiltonian for
the state $[I^{(3)}_{\rm max} \times j]^{I^{(4)}}$. Suppose that the lowest
 spin $I^{(4)}$  state is not degenerate.
Then a certain state which produces the {\it same} energy as the lowest
spin $I^{(4)}$ state will have the same
 wavefunction.  
In Table I most of the energies obtained by diagonalizing $H_{J_{\rm max}}$ for
$n=4$ are   close to  the matrix element for the pure configuration of the
$[I_{\rm max}^{(3)}\times j]^{I^{(4)}}$ state (also close to $E_{I_{\rm
max}^{(3)}}$). Overlaps of states having other ``non-integer" eigenvalues
near $-2.25$ for $n=4$ with those given by the $[I_{\rm max}^{(3)}\times
j]^{I^{(4)}}$ are close to 1, except three cases (two of them can be approximated by
other  three-particle clusters with $I^{(3)} \sim I^{(3)}_{\rm max}$
(but $I^{(3)} \neq  I^{(3)}_{\rm max}$)  coupled to a single particle).

We have calculated all overlaps between states
of $n=4$ which have energies close to the peaks
and those of simple wavefunctions 
obtained by coupling  a single particle to a non-zero energy cluster with
$I^{(3)} \sim I^{(3)}_{\rm max}$ of three fermions.
These show a similar situation as Table I.
Therefore, we conclude that those  stable   ``non-integer"
 eigenvalues of $n=4$ with
$H = H_{J_{\rm max}}$ in Fig. 1 are given to a high precision by a
three-particle cluster (nonzero energy) coupled to a single-$j$ particle.

One may ask which picture is more relevant to the states of $n=4$ with
eigenvalues close to integers; one in which a three-particle cluster (nonzero
energy) is coupled  to a single-$j$ particle, $[I^{(3)}\times 
j]^{I^{(4)}}$ with
$I^{(3)}\sim I^{(3)}_{\rm max}$, as proposed in this paper,
or one in which four particles are coupled
pairwise with one or two spins being $J_{\rm max}$, 
 as proposed in Ref. \cite{nucl-th/0305095}? 

First we note that for $I^{(4)} \ge 53$ only a single state is possible and these two
pictures are therefore equivalent and exact; for $I^{(4)}=0$ (or 3)  the number of
states is the largest integer not exceeding $(2j+3)/6$ (or $(2j-3)/6$) which is
larger than 1 in most cases \cite{dimension} but there is only one state which
gives a non-zero eigenvalue for the ${J_{\rm max}}$ pairing interaction
\cite{nucl-th/0305095}. Also in this case the two pictures are therefore
equivalent and exact.

For states with $I^{(4)} < J_{\rm max}$  and
energy close to $-2.0$, it was proven in Ref. 
\cite{nucl-th/0305095} that a description by using two pairs with   two spins
being $J_{\rm max}$ is very good. For these states one may ask whether a description
by using a single-$j$ particle coupled to one of three-particle clusters with
$I^{(3)} \sim I_{\rm max}^{(3)}$
  is also relevant. To see
whether or not this is true, we calculate  the overlaps of 
the  states of four fermions 
with $E_{I^{(4)}, J_{\rm max}(j)}^{(4)} \sim -2.0$
and $I^{(4)} < J_{\rm max}$ 
which were obtained by the shell model diagonalization with 
{\it all} possible three-particle clusters 
coupled with a single-$j$ particle.  These overlaps  are between
around 0.6-0.8. Thus the picture using  a single-$j$ particle coupled to the 
three-particle cluster for states with $I^{(4)} < J_{\rm max}$  and 
energy close to $-2.0$  is not appropriate.

Then, how does the picture of two pairs with spin $J=J_{\rm max}$ work as $I^{(4)}$
increases? We calculate the overlaps of states with energy around $-2$. We see
that for  $I^{(4)}=42$ (=$I^{(4)}_{\rm max} - 14$) the overlap is still 0.9962, showing
that the  pair picture is still very good.

The next question is related to the states
of four fermions with energies near $-1$. There are  
about 100 states for $n=4$ and $j=31/2$. Both pictures can
give eigenvalues at $-1$. 
The number of states with $E\sim -1$ is not unique. As was shown in Ref. 
\cite{nucl-th/0305095}, for states with small $I^{(4)}$, the number of states with 
$E_{I^{(4)}, J_{\rm max}(j)}^{(4)}\sim -1$
states   is the largest integer not exceeding $I^{(4)}/2$, which is
larger than 1 except for $I^{(4)}=0,$ 2 and 3. Because these
eigenvalues are very close to but {\it not
exactly} $-1$,  the mixing of these configurations 
can be  large. Coupling two pairs, one with spin $J_{\rm max}$ and the  other spin $J'
\neq J_{\rm max}$ picture gives a very good classification of states, but not the exact
wavefunctions. On the other hand, the picture using a three-particle cluster of
nonzero energy coupled with a single-$j$ particle provides us with better
wavefunction than the pair picture, but it does not provide  us the number of
states with $E_{I^{(4)}, J_{\rm max}(j)}^{(4)} \sim -1$. These two pictures
are therefore complementary in describing the
states for $n=4$ with $E_{I^{(4)}, J_{\rm max}(j)}^{(4)} \sim -1$.

\section{States of five particles and those
of six particles with $H=H_{J_{\rm max}}$}

In this section, we proceed to 
 more particle systems.  Although we did not find simple 
descriptions for them,  we 
are able to find some relations
between states of different $n$ systems with an attractive 
$J_{\rm max}$ pairing interaction. 

The picture using clusters of ${\cal N}$ (${\cal N}<n$) particles coupled to
$(n-{\cal N})$ single-$j$ is also found in states
of systems with $n>4$. We study 
in this section the $j=19/2$ shell for both $n=5$ and 6.
The cases with larger 
$j$-shells yield a similar picture with  higher accuracy.

Asymptotic integers appear in the eigenvalues $E_{I^{(5)}, J_{\rm max}(j)}^{(5)}$ 
  when $I^{(5)}$ is not very 
large. They are either zero, or very close to $-1$ and $-2$. The 
number of states for $I^{(5)}=\frac{1}{2}$ is three, among which there is one with 
zero eigenvalue, one with eigenvalues $\sim -1$ (within a precision of 0.01)
and one with eigenvalue $\sim -2$ (within a precision of 0.01). The number of
states for $I^{(5)}=\frac{3}{2}$ is seven, among which there are two with zero
eigenvalues, three with eigenvalues $\sim -1$ (within a precision of 0.01) and
two with eigenvalues $\sim -2$ (within a precision of 0.01). A similar
situation holds for larger $I^{(5)}$ states except that
 eigenvalues $\sim (E_{I^{(3)}, J_{\rm max}(j)}^{(3)}-1)$
($I^{(3)}\sim I^{(3)}_{{\rm max}}$)
or $E_{I^{(3)}, J_{\rm max}(j)}^{(3)}$ appear.
Corresponding to each  three-body cluster with 
$I^{(3)}\sim I^{(3)}_{{\rm max}}$, the minimum of $I$ which 
gives   ``non-integer" eigenvalues
 $E_{I^{(5)}, J_{\rm max}(j)}^{(5)}
\sim (E_{I^{(3)}, J_{\rm max}(j)}^{(3)}-1)$
and $ E_{I^{(5)}, J_{\rm max}(j)}^{(5)}
\sim E_{I^{(3)}, J_{\rm max}(j)}^{(3)}$ 
 is  ${I}^{(3)}-(2j-1)$ and   ${I}^{(3)}-(2j-3)$,
respectively.
For examples, $E_{I^{(5)}, J_{\rm max}(j)}^{(5)}  \sim
E_{I^{(3)}_{\rm max}-5, J_{\rm max}(j)}^{(3)} -1 =  
\frac{6469}{3410}=-$1.89707  appears in states with 
$I^{(5)} \ge I^{(3)}_{\rm max}-5 -(2j-1)  = j-7= \frac{5}{2}$; 
 $E_{I^{(5)}, J_{\rm max}(j)}^{(5)}  \sim
E_{I^{(3)}_{\rm max}-5, J_{\rm max}(j)}^{(3)} =  
\frac{3059}{3410}=-$0.89707  appears in states with 
$I^{(5)} \ge I^{(3)}_{\rm max}-5 -(2j-3)  = j-5= \frac{9}{2}$, etc. 
The non-zero eigenvalues for $n=5$ are 
 equal to or concentrated around 0, $-1$, $-2$, 
$\sim E_{I^{(3)}, J_{\rm max}(j)}^{(3)} $ and
$\sim (E_{I^{(3)}, J_{\rm max}(j)}^{(3)}-1) $  with
$I^{(3)} \sim I^{(3)}_{\rm max}$. 
The above regularities survive
unless $I^{(5)} \sim I^{n=5}_{\rm max}$.

Now let us look at the case with $n=6$ in the same shell $j=19/2$.
Below we consider the
case with $I^{(6)}=0$ as an example because other low $I^{(6)}$
states behave similarly. There 
are ten states with $I^{(6)}=0$.  Among them there are two
with zero eigenvalues. 
Non-zero eigenvalues are more complicated than systems with smaller $n$,
because $n=6$ can be divided into more sets of clusters. We first divide 
$n=6$ into two clusters with $n=3$ and $I^{(3)} 
\sim I_{\rm max}^{(3)}$ for each three-body cluster.  
Then we obtain eigenstates with  eigenvalues: $-$4.54286, $-$3.31055,
$-$1.23269, $-$2.42027, $-$1.75573, and $-$2.10284, which are 
approximately equal to  twice those of
$E_{I^{(3)}, J_{\rm max}(j)}^{(3)}$ ($I^{(3)} 
\sim I^{(3)}_{\rm max}$): $\sim
-\frac{159}{35}, -\frac{182}{55}, -\frac{95}{77},
-\frac{1645}{682}, -\frac{3059}{1705}, -\frac{20727}{9889}$, respectively.
We can also divide $6$ into three two-body pairs. Here we take  
$I^{(2)}=J_{\rm max}$ for these three pairs
which lead  to an eigenvalue very close 
to $-3$ ($-$3.01537).
Besides these eigenvalues, there are one  eigenvalues which 
are close to $-1.0$.

We did not succeed in setting up a simple scenario of the  
distribution for all eigenvalues of systems  with $n=5$ or 6
and  $H=H_{J_{\rm max}}$. 
This is partly because the number of states
for each $I^{(n)}$ is not analytically 
known. The number of combinations for different  clusters
is also much larger than the cases with $n=3$ and $n=4$.

Based on these relations we suggest that   the low-lying states  
of each $I^{(n)}$ of fermions (bosons) in a single-$j$ shell (with  spin $l$)
interacting by an attractive $H=H_{J_{\rm max}}$ favor a cluster structure,
where each cluster has a maximum (or close to maximum) angular momentum. The 
coupling between the constituent clusters (including pairs and spectators) 
 are very weak and negligible, therefore we
can obtain both their approximate wave functions and  eigenvalues, which are
simple summation of those of the clusters.

In Fig. 2, we showed
the distribution of all non-zero eigenvalues for systems 
with  $n$ ranging from 2 to 6. It is easy to 
notice that the eigenvalues are concentrated around 
some values for $n=2$ to 5.
This pattern becomes less striking for $n=6$.

\section{Discussion and summary}

In this paper, we first show that a system of three fermions in a single-$j$
shell in the presence of $H=H_J$  is solvable. We prove that there is at most
one state with a non-zero eigenvalue for each $I^{(3)}$.
We can analytically  construct
both the eigenvalues and corresponding wave functions.  A similar remark
applies to three bosons with spin $l$ in the presence of $H_J$. On the basis of
the above results for $n=3$ a series of new sum rules of six-$j$ symbols can be
found.

We show  that the eigenvalues of three fermions in a single-$j$ shell
with $H=H_{J_{\rm max}}$ are very close to 0 or $-1$ unless
$I^{(3)} \sim I^{(3)}_{\rm max}=3j-3$. This kind of situation
is very similar to the case of $n=4$, 
as studied in  Ref. \cite{nucl-th/0305095}. 

We also find that the ``non-integer" eigenvalues of
 $I^{(3)}\sim I^{(3)}_{\rm max}$ for 
$n=3$ appear as ``non-integer" eigenvalues for $n=4$ when
$I^{(4)}$ is around or larger than
$J_{\rm max}$. The overlaps between the wavefunction of these
``non-integer" eigenvalues of $n=4$ and that
of $I^{(3)} \sim I^{(3)}_{\rm max}$ state 
coupled to a single-$j$ particle  is very close to 1. This 
finding allows us to construct approximately
the states of $n=4$ by using results 
of $n=3$ as we have shown. We confirmed that this is also true for  five and
six fermions in a single-$j$ shell in the presence of $J_{\rm max}$ pairing
interaction. Bosons with spin $l$ exhibit a similar pattern.
Similar regularity was found for $n=5$ and 6, although we 
did not succeed in setting up a simple rule for all states.

The  relations between
 $E_{I^{(2)}, J_{\rm max}(j)}^{(2)}$,  $E_{I^{(3)}, J_{\rm max}(j)}^{(3)}$ 
  $E_{I^{(4)}, J_{\rm max}(j)}^{(4)}$ $\cdots $ 
 indicate the  following pattern:
the attractive $J_{\rm max}$ pairing interaction
 favors  clusters (including pairs and spectators), 
 where the angular momentum of each cluster is close to the
 maximum.  One thus explains  the ``integer" eigenvalues 
and ``non-integer" eigenvalues
proposed in Ref. \cite{nucl-th/0305095} by using a picture of the 
clusters for fermions in a single-$j$ shell or bosons with spin $l$.

As is well known, the existence of degeneracy indicates that the Hamiltonian
has a certain symmetry.
The degeneracy for the $J_{\rm max}$ pairing interaction,
 however, is not exact.
It would be interesting to explore  the broken
symmetry hidden in the $J_{\rm max}$ pairing interaction discussed
in this paper.  It would be also interesting to discuss 
the modification of the $J_{\rm max}$ pairing interaction in order to
recover the exact degeneracy.

{\bf Acknowledgement:}
We wish to extend our special thanks to Prof. I. Talmi for his
 valuable comments  concerning the sum rules of six-$j$ symbols. 
We also thank  Dr. N. Yoshinaga for discussions in the early stage  
of this work,  and Drs. I. Talmi and O. Scholten for their 
reading of this manuscript.

\newpage

{ Table I ~  ~ The lowest eigenvalues   of the $I^{(4)}$ states 
 in a single-$j$ ($j=31/2$) shell with 
$I^{(4)}$ between 28 to $I^{(4)}_{\rm max}=56$. When $I^{(4)}$ is smaller 
than 48 there is no eigenvalue lower than $-2$.
The eigenvalue of the $I^{(3)}_{\rm max}$ state with three fermions
in the same single-$j$ shell is $-\frac{267}{118}$=$-2.26271186440677966$.
The column ``(SM)" is obtained by a shell model
diagonalization, and the  column ``${\cal E}_I$" is matrix element
of $H_{J_{\rm max}}$ for 
 the  state constructed by 
  three-fermion  with $I^{(3)} = I^{(3)}_{\rm max}$ coupled to  a  
spectator. The column ``error" presents the difference between 
${E}_I$  and ${\cal E}_I$ (two effective digits). 
The column ``overlap" is the overlap between
the lowest eigenstates of $n=4$ and the states obtained by
coupling single fermion $a_j^{\dag}$ to the $I^{(3)}_{\rm max}$ state. 
Italic font is used for three cases for which the overlap
is not close to 1. We note that the
case of $I^{(4)}=50$ (52) can be approximated rather accurately ($10^{-3}$) as 
  a three-particle cluster with $I^{(3)}=I^{(3)}_{\rm max} -8 $
 ($I^{(3)}_{\rm max} -6 $)  coupled to a single-$j$ spectator. }

\vspace{0.2in}

\begin{tabular}{c|cccc} \hline \hline
 $I$ & ${E}_I$ (SM) ~  & ${\cal E}_I$ (coupled)& ``error" & overlap  \\ \hline
28 & -2.26271186440689  & -2.262711864406782 & $1.1\times 10^{-13}$ & 1.000000000000000 \\
29 & -2.26271186440682  & -2.262711864406777 & $4\times 10^{-14}$   & 1.000000000000000 \\
30 & -2.26271186440678  & -2.262711864406780 & $1 \times 10^{-14}$  & 1.000000000000000    \\
31 & -2.26271186440669  & -2.262711864406782 & $0.9\times 10^{-13}$ & 0.999999999999999    \\
32 & -2.26271186440692  & -2.262711864406805 & 1.1$\times 10^{-13}$ & 0.999999999999659    \\
33 & -2.26271186440700  & -2.262711864406981 & $2\times 10^{-14}$   & 1.000000000000001    \\
34 & -2.26271186442899  & -2.262711864409884 & $1.9\times 10^{-11}$ & 0.999999999963606    \\
35 & -2.26271186442233  & -2.262711864422405 & $8\times 10^{-14}$   & 0.999999999999996    \\
36 & -2.26271186573172  & -2.262711864593695 & $1.1\times 10^{-9}$  & 0.999999997833653    \\
37 & -2.26271186512903  & -2.262711865128374 & $6.6\times 10^{-13}$ & 0.999999999999758    \\
38 & -2.26271191546116  & -2.262711871692375 & $4.3\times 10^{-8}$  & 0.999999916591887    \\
39 & -2.26271188689249  & -2.262711886864667 & $2.8\times 10^{-11}$ & 0.999999999988818    \\
40 & -2.26271325181426  & -2.262712064607266 & $1.2\times 10^{-6}$  & 0.999997726566892    \\
41 & -2.26271236805292  & -2.262712367094411 & $9.6\times 10^{-10}$ & 0.999999999598199    \\
42 & -2.26274016611845  & -2.262715960012236 & $2.4\times 10^{-5}$  & 0.999952666087478    \\
43 & -2.26272031287460  & -2.262720286322401 & $2.7\times 10^{-8}$  & 0.999999987392690    \\
44 & -2.26317530567842  & -2.262776481261782 & $4.0\times 10^{-4}$  & 0.999151747579904    \\
45 & -2.26282037299297  & -2.262819747102017 & $6.2\times 10^{-7}$  & 0.999999632523561    \\
46 & -2.26963309159052  & -2.263514816015588 & $6.1\times 10^{-3}$  & 0.982828211942919    \\
47 & -2.26378385186917  & -2.263772302947436 & $1.2\times 10^{-5}$  & 0.999992036003522    \\
48 & -2.34719850307215  & -2.270625142453812 & $7.7\times 10^{-2}$  & {\it 0.780582505446094}    \\
49 & -2.27068252318197  & -2.270571840272616 & $1.1\times 10^{-4}$  & 0.999929221753443    \\
50 & -2.57872583562800  & -2.323429204525185 & $2.6\times 10^{-1}$  & {\it 0.706859839896674}    \\
51 & -2.30488200470359  & -2.304882004703592 & 0    & 1.000000000000000 \\
52 & -2.89017281282010  & -2.592166600952603 & $3.0\times 10^{-1}$  & {\it 0.873170713095796}    \\
53 & -2.41926851025870  & -2.419268510258698 & 0 & 1.000000000000000 \\
54 & -3.24511394047522  & -3.245113940475225 & 0 & 1.000000000000000 \\
56 & -3.66369313113292  & -3.663693131132918 & 0 & 1.000000000000000 \\
 \hline \hline
\end{tabular}

\vspace{0.3in}

\newpage

\begin{center}

Appendix {\bf A}  New sum rules of six-$j$ symbols

\end{center}

The solution of $H_J$ for $n=3$  gives new sum rules.
The procedure to obtain these sum rules is straightforward. As is
well known,
the summation of all eigenvalues with a fixed $I$ is equal to
$\frac{n(n-1)}{2}$ times the number of $I$ states, where $n$ is
the particle number.
For $n=3$, the number of states can be empirically expressed in a compact
formula   \cite{dimension}.

In Ref. \cite{nucl-th/0305095} we applied this idea and obtained  that
\begin{eqnarray}
 \sum_{{\rm even~}J} (2J+1)
    \left\{ \begin{array}{ccc}
    j    & j  & J \\
    j    & j  & J  \end{array} \right\} = \frac{3}{2}
 \left[ \frac{2j+3}{6} \right] -\frac{2j+1}{4}
  =
    \left\{ \begin{array}{cl}
    \frac{1}{2}    & {\rm if} ~ 2j=3k,    \\
    0              & {\rm if} ~ 2j =3k+1,    \\
- \frac{1}{2}    & {\rm if} ~ 2j=3k+2,   \end{array} \right.
\label{old1}
\end {eqnarray}
where $j$ is a half integer, and $\left[ x   \right]$ means to take the
largest integer  not exceeding $x$.
we   derive a similar sum rule using the
$I=0$ states of four bosons with spin $l$:
\begin{eqnarray}
&& \sum_{{\rm even~}J} (2J+1)
    \left\{ \begin{array}{ccc}
    l    & l  & J \\
    l    & l  & J  \end{array} \right\} = \frac{3}{2}
 \left[ \frac{l}{3} \right] + 1 -\frac{l}{2}  =
    \left\{ \begin{array}{cl}
    1    & {\rm if} ~ l=3k,    \\
    \frac{1}{2}               & {\rm if} ~ l=3k+1,    \\
    0    & {\rm if} ~ l =3k+2,   \end{array} \right.
\label{old2}
\end{eqnarray}

Below we give other sum rules of six-$j$ symbols. For  a half integer $j$,
\begin{eqnarray}
\sum_{J={\rm even}} 2(2J+1)
    \left\{ \begin{array}{ccc}
    j    & I  & J \\
    j    & j  & J  \end{array} \right\}
=    \left\{ \begin{array}{ll}
3 \left[ \frac{2I+3}{6} \right] - I - \frac{1}{2}  &  {\rm if ~} I \le j; \\
3 \left[ \frac{3j-3-I}{6} \right] + 3\delta_I^j- \left[ \frac{3j+1-I}{2}
 \right]  & {\rm if ~} I \ge j ~.
\end{array} \right.
\label{new1}
\end{eqnarray}
where
\begin{eqnarray}
&& \delta_I^j   =  \left\{
\begin{array}{ll}
0      &  {\rm if}~ (3j-3-I) {\rm ~ mod}~ 6 = 1   \\
1       &  {\rm otherwise} ~.  \nonumber
\end{array}  \right.
\end{eqnarray}

For integer $l$, we obtain similar
sum rules given as follows. For  $I \le l$ ($l$ is an integer),
\begin{eqnarray}
\sum_{J={\rm even}} 2(2J+1)
    \left\{ \begin{array}{ccc}
    l    & I  & J \\
    l    & l  & J  \end{array} \right\}
=     \left\{ \begin{array}{ll}
3   \left[ \frac{I}{3} \right]  -  I
+1 + (-)^{I+l}     &   {\rm if ~} I \le l; \\
3 \left[ \frac{3l-I}{6} \right] + 3 \delta_I^l
 - \left[ \frac{ 3l-I+2}{2} \right]   & {\rm if ~}  I \ge l ~.
\end{array} \right.
\label{new2}
\end{eqnarray}
where
\begin{eqnarray}
&& \delta_I^l   =  \left\{
\begin{array}{ll}
0      &  {\rm if}~ (3l-I)  {\rm ~ mod}~   6 = 1   \\
1       &  {\rm otherwise} ~.  \nonumber
\end{array}  \right.
\end{eqnarray}

It is noted  that the sum rules (\ref{old1}) and (\ref{old2}) are
special cases of the  sum rules (\ref{new1}) and (\ref{new2}).

Starting from Eq.(10.14) of Ref. \cite{Talmi-II} for
$J'=J'', J_1=J_3=j, J_2=J_4=j'$, we multiply $(2J+1)$ and sum over $J'$.
Using  Eq. (10.13), we  obtain 
\begin{eqnarray}
&& \sum_{J} (2J+1)
    \left\{ \begin{array}{ccc}
    j    & j'  & J \\
    j    & j'  & J  \end{array} \right\}
=0 ~ ;  \nonumber \\
&& \sum_{J} (2J+1)
    \left\{ \begin{array}{ccc}
    l    & l'  & J \\
    l    & l'  & J  \end{array} \right\}
=1 ~.  
\end{eqnarray}
Similarly, we  obtain 
\begin{eqnarray}
&& \sum_{J}  (2J+1)
    \left\{ \begin{array}{ccc}
    l    & j  & J \\
    l    & j  & J  \end{array} \right\}
  =  \left\{
\begin{array}{ll}
-1      &  {\rm if}~ l < j ~,    \\
0       &  {\rm otherwise} ~.  
\end{array}  \right. 
\label{new5}
\end{eqnarray}
and 
\begin{equation}            
\sum_J 2 (2J+1)
    \left\{ \begin{array}{ccc}
    j    & I  & J \\
    j    & j  & J  \end{array} \right\}
  =  \left\{
\begin{array}{ll}
0      &  {\rm if}~ I \le j ~,    \\
(-)^{I-j} -1       &  {\rm if}~ I \ge j ~.  \nonumber
\end{array}  \right.
      \label{j-sum1}
\end{equation}
Using Eqs. (\ref{j-sum1}) and
 (\ref{new1}), we obtain 
\begin{equation}
\sum_{J={\rm odd}} 2 (2J+1)
    \left\{ \begin{array}{ccc}
    j    & I  & J \\
    j    & j  & J  \end{array} \right\}
= I+ \frac{ 1}{2} -
3 \left[ \frac{2j+3}{6} \right]
  =  \left\{
\begin{array}{ll}
-1      &  {\rm if}~  2j = 3 k~,    \\
0      &  {\rm if}~ 2j = 3k+1 ~,    \\
1       &  {\rm if}~ 2j = 3k+2 ~.  \nonumber
\end{array}  \right.
\end{equation}
and
for $I \ge j$,
\begin{equation}
\sum_{J={\rm odd}} 2 (2J+1)
    \left\{ \begin{array}{ccc}
    j    & I  & J \\
    j    & j  & J  \end{array} \right\}
=(-)^{I-j} -1 - 3 \left[ \frac{3j-3-I}{6} \right]
- 3\delta_I^j + \left[ \frac{3j+1-I}{2} \right]
\end{equation}

Similar to Eq. (\ref{j-sum1}), we  obtain  
\begin{equation}
\sum_{J} 2(2J+1)
    \left\{ \begin{array}{ccc}
    l    & I  & J \\
    l    & l  & J  \end{array} \right\}
  =  \left\{
\begin{array}{ll}
2 (-)^{I+l} ~       &  {\rm if}~  I \le l~,    \\
1+(-)^{I+l} ~       &  {\rm if}~ I \ge l ~.  \nonumber
\end{array}  \right.
\label{l-sum1}
\end{equation}
Using Eqs. (\ref{l-sum1}) and
(\ref{new2}), we obtain that
\begin{equation}
\sum_{J={\rm odd}} 2(2J+1)
    \left\{ \begin{array}{ccc}
    l    & I  & J \\
    l    & l  & J  \end{array} \right\}
  =  \left\{
\begin{array}{ll}
 (-)^{I+l} + I - 3   \left[ \frac{I}{3} \right]-1
        &  {\rm if}~  I \le l~ ;   \\
(-)^{I+l} - 3 \left[ \frac{3l-I}{6} \right] - 3 \delta_I^l
- \left[ \frac{3l-I}{2} \right] + 2~
       &  {\rm if}~ I \ge l ~.
\end{array} \right.
       \nonumber
\end{equation}

\newpage

Caption:

\vspace{0.3in}

Fig. 1 ~~
Detailed distribution  of all non-zero eigenvalues for $n=4$.
The inset in Fig. 1(b) is rescaled  to distinguish a few
exceptional cases which energies  are  not close to those of $n=3$.
(a), (b) and (c) corresponds to different range of
$|E_{I^{(4)}, J_{\rm max}(j)}^{(4)}|$.
We see that the eigenvalues for $n=4$ are ``clustered"
at those of $n=3$ with few exceptions.

\vspace{0.3in}

Fig. 2 ~~Distribution of non-zero eigenvalues
$|E_{I^{(n)}, J_{\rm max}(j)}^{(n)}|$
for systems with $n$ ranging from 2 to 6 and $j=19/2$.
One sees that the non-zero eigenvalues are highly concentrated.
The concentration of eigenvalues for the case of $n=6$ is less
striking. The distribution is plotted using 
the number of counts for each
$|E_{I^{(n)}, J_{\rm max}(j)}^{(n)}|$  (with the
step length being 0.01) divided 
by the total number of non-zero eigenvalues.
For  $H=H_{J_{\rm max}}$ with
$j=19/2$, the number of  non-zero eigenvalues 
 is 1, 17, 122, 472, 1224
(in comparison with the number of
the shell model space: 10,  45, 177, 521, 1242)
for $n=2$, 3, $\cdots$, 6, respectively.

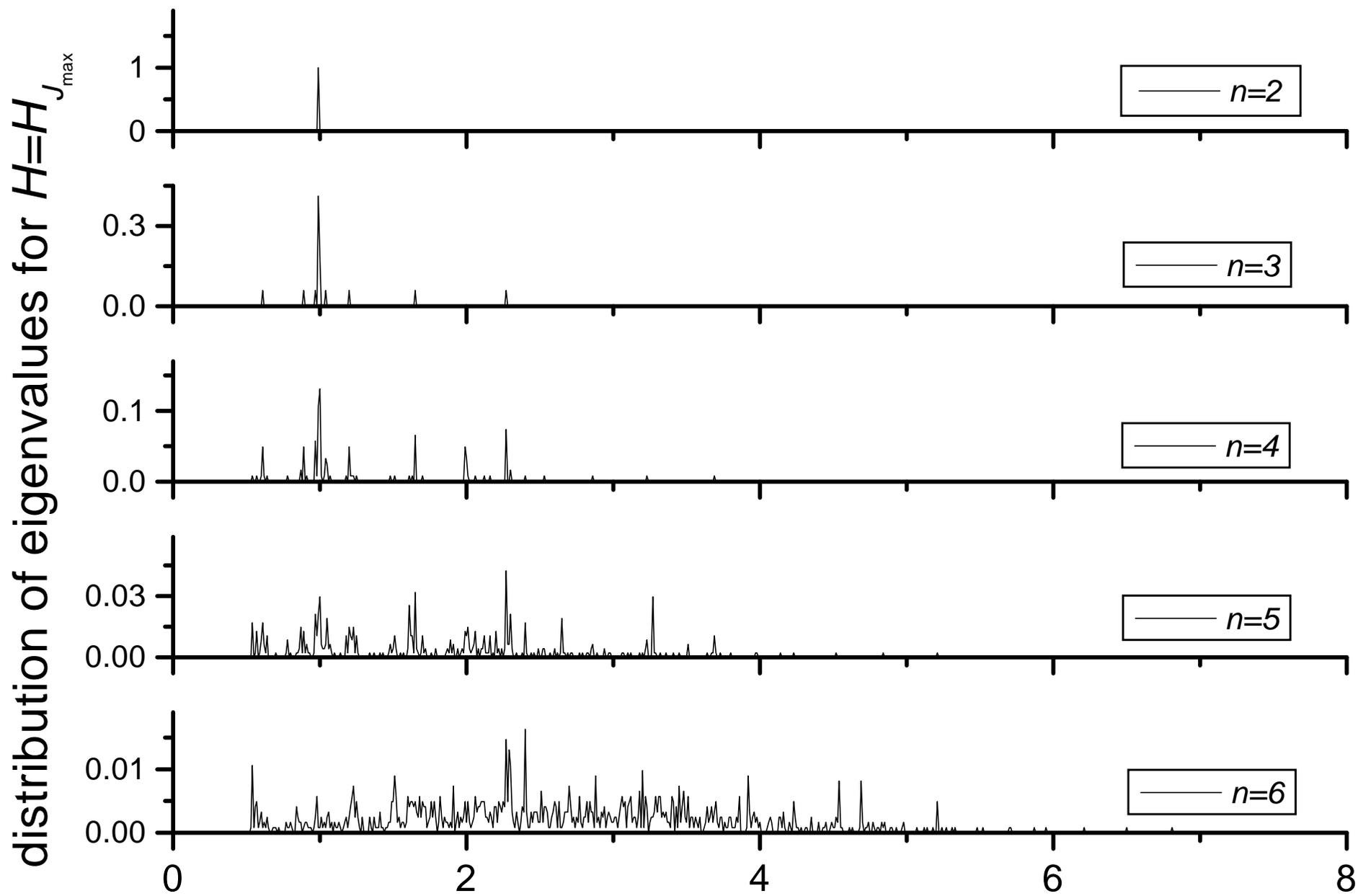
\end{document}